# Designable hybrid sonic crystals for transportation and division of acoustic images

Zhaojian He[1a)], Ke Deng[1b)], Heping Zhao[1], and Xiaochun Li[2, 3]

1. Department of Physics, Jishou University, Jishou 416000, Hunan, China

2. School of Physics Science and Technology, Xinjiang University, China

3. Department of Physics and Electronics, Central South University, China

**Abstract:**

Conventional sonic crystal (SC) devices designed for acoustic imaging can focus acoustic waves from an input source into only one image but not multi-images. Furthermore the output position of formed image cannot be designed at will. In this paper, we propose the hybrid SC imaging devices to achieve multi-images from one-source-input along with the designable image-positions. The proposed hybrid devices can image acoustic waves radiated both from point source and Gaussian beam, which different from conventional SC imaging devices that only applies to point source. These multi-functional but still simple and easy-to-fabricate devices are believed to find extensive applications, particularly in ultrasonic photography and compact acoustic imaging.

Author to whom correspondence should be addressed to: [a)] hezj@jsu.eud.cn, and [b)]dengke@jsu.edu.cn.

Sonic crystals (SCs) efficiently guide the propagation of acoustic waves, and thus provide an exhilarating tool for arbitrary manipulation of acoustic waves. Suitable designing of SCs can induce many promising and exciting acoustic applications. For involving in the applications based on SCs, the waveguides and focusing devices are very popular and have been extensively studied [1-10]. Concerning to acoustic waveguides, there are generally two methods of construction: one is to employ the self-collimating effect of SCs [4, 5]; the other is to introduce defects in the band gaps of SCs [1-3]. While for the imaging devices, the negative refraction and direct canalization effects are mostly used to achieve the goal [6-10].

However, as far as we know, by now most existing SC imaging devices can focus the acoustic waves from an input source into only one image but not multi-images [6-10]. Moreover, the formed image is always confined in a fixed place near the devices, which means that the output imaging-position cannot be designable. These two flaws in SC imaging may cause great inconvience in the ultrasonic photography and acoustic compact devices. To remedy these flaws, in this paper we propose a hybrid SC that consists of two different simple SCs. One of the simple SCs acts as a band gap waveguide, and the other works in the negative refraction effect. With the combined effects of wave-guiding and negative refraction, one-source-input to multi-images as well as designable image-position can be achieved by the proposed hybrid SC devices. Furthermore, this hybrid imaging device can focus the acoustic waves radiated both from point source and Gaussian beam, which gains more applicability than the normal negative refraction SCs that only applies to point source. At the same, compared to the gradient-index SCs that always focus the Gaussian beam [11-13], hybrid SCs are intrinsically much simpler and then much easier to fabricate. Thus devices based on the hybrid SC are much easier to realize.

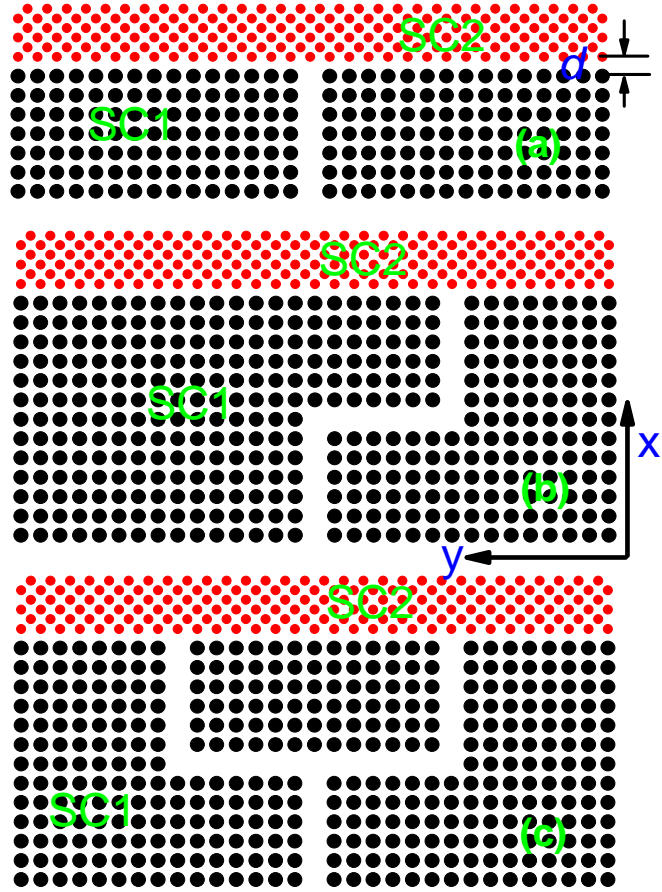

**Figure 1.** Structures for the acoustic images transportation and division: (a) straight-line transportation structure, (b) bending transportation structure, and (c) division structure. Sonic crystal 1 (SC1) is aimed to design as waveguide created by removing one row of cylinders. Sonic crystal 2 (SC2) is aimed to design as negative-refraction imaging device.

To be more specific, let us consider two-dimensional hybrid sonic crystals that can realize the waves functions described above, as schematically illustrated in Fig. 1, which are merged by two different simple SCs. The SCs consist of water cylinders immersed in mecury with suqare array. The elastic parameters of the involved materials are as follows [14]: $\rho$=13.5 g/cm$^3$, $c_L$=1.45 km/s for mercury; $\rho$=1.0 g/cm$^3$, $c_L$=1.49 km/s for water, where $\rho$ and $c_L$ are the density and the longitudinal sound velocity respectively. The first simple sonic crystal (SC1, marked with black solid cycles) sets along ΓX direction and the cylinder radius is 0.3$a$. Here, $a$ is the unitary constatnt. The second sonic crystal (SC2, marked with red solid cycles) sets along ΓM direction and the cylinder radius is 0.28$a$. The period for the two simple SCs is 1.0$a$, and the distance $d$ of the two simple SCs edge cylinders, as shown in Fig. 1(a), is 1.0$a$. Throughout this paper, we adopt the rigors multiple scattering theory method [15] to calculate the band structures and field distribution.

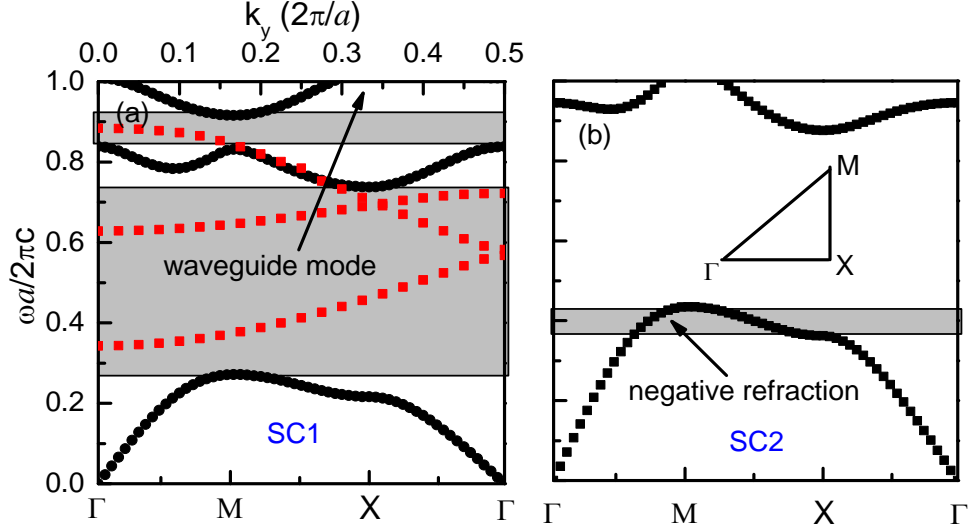

**Figure 2. Band structures of the sonic crystals studied. (a) Black dots are the band structures of the sonic crystal 1. Shade area indicates the band gaps. Red dots are the band structures of the waveguides. $K_y$ is the wave vector along the waveguides propagation direction, c is the sound velocity in mercury. (b) Band structures of sonic crystal 2. Shade area indicates the negative refraction area.**

In order to clearly analyze the working mechanism of the hybrid sonic crystals, we study the band structures of the two simple SCs. The results are shown in Fig. 2. The black dots represent the band structures of the two sonic crystals (left panel for SC1 and right panel for SC2), and red dots represent the band structures of the waveguide created by removing one row of cylinders in SC1. It can be observed that, the first band gap of SC1 [marked by grey area in Fig2. (a)] is so wider that the first negative refraction-band of SC2 [marked by grey area in Fig2. (b)], of which the frequency ranges from 0.36$\omega a/2\pi c$ to 0.44$\omega a/2\pi c$, is entirely located within it. Further, there are several waveguide bands in the band gaps of SC1, and the first band of these waveguide modes, of which the frequency ranges from 0.34$\omega a/2\pi c$ to 0.56$\omega a/2\pi c$, is fully across the negative refraction band of SC2. Therefore, for the hybrid SC structures in Fig.1, any incident acoustic waves with frequency within the negative fraction band of SC2 can be guided across the SC1 component (the black solid cycles) by the waveguide modes, and then be focused into images by negative refraction when encountered with the SC2 component (the red solid cycles). As a result, by designing different functional waveguides in the SC1 component, e.g. the transportable waveguides and branching waveguides, images of different functionalities, such as the one-source-input to multi-images and the controllable image-position, can be realized through negative fraction of the SC2 component as illustrated in Fig 1. At the same time, as the flow of the incident wave in our hybrid SC is controlled by the wave guiding functions of SC1 which are independent of the input radiated sources, hence these functional imaging process can be achieved for both the point sources and the Gaussian sources.

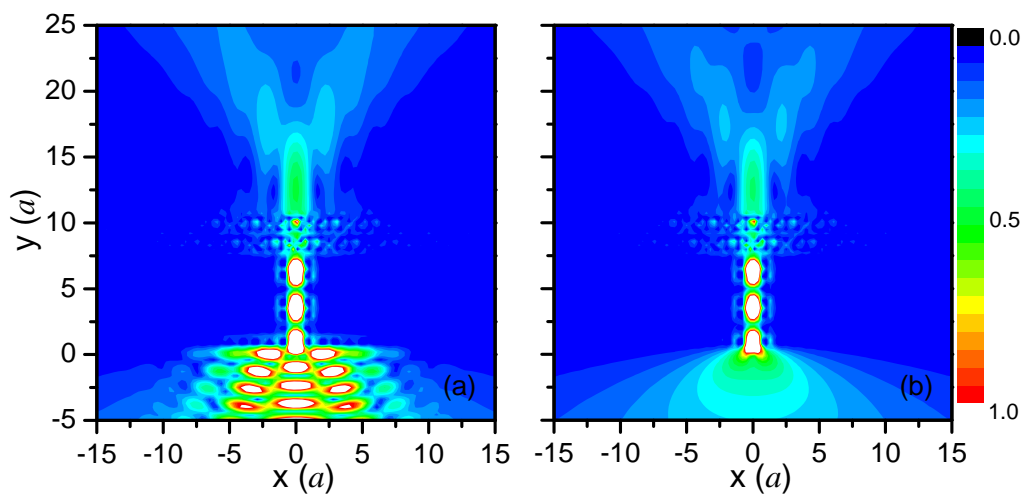

**Figure 3. Imaging pressure field of the straight-line transporting hybrid sonic crystal [structure of Fig. 1(a)] for Gaussian source incidence (a), and point source incidence (b) at frequency 0.38ω$a$/2π$c$.**

To verify the above analysis, in Fig. 3 we demonstrate a straight-line transportation of the acoustic images at frequency 0.38$a$/2π$c$ by the hybrid sonic crystal [structure of Fig. 1(a)]. We can see that, for both the Gaussian incident source and point incident source, through the hybrid sonic crystal an acoustic image is clearly formed in the other side of the hybrid SC. It clearly manifests here that this unique imaging device is independent to the incident source as expected above; moreover, as a device to focus Gaussian source the involved hybrid sonic crystal is much easier to be constructed in contrast to the gradient-index SCs. Furthermore, by tuning the length of the waveguide in SC1, the distance between source and image can be easily controlled at will.

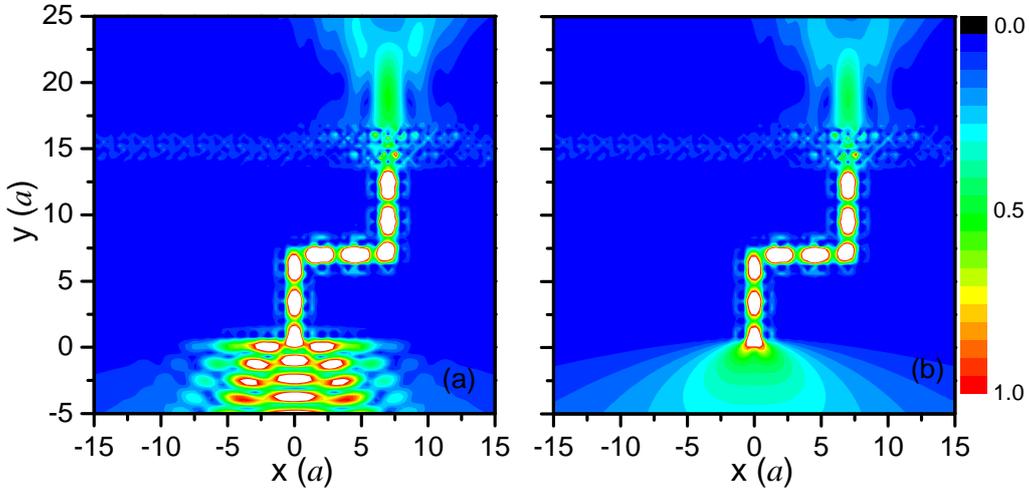

**Figure 4. Imaging pressure field of the bending transporting hybrid sonic crystal [structure of Fig. 1(b)] for Gaussian source incidence (a), and point source incidence (b) at frequency 0.38$\omega a/2\pi c$.**

It shows in Fig. 3 that acoustic waves radiated from the input sources can be transported firstly by the waveguide in SC1, and then be focused onto an image by the negative refraction in SC2. The source-image-distance is completely determined by the transporting path in SC1. Actually, transporting path of acoustic waves can be controlled *at will* through tuning the waveguide property in SC1, and hence the image position of our hybrid SC can be designed at will. In Fig. 4, we exemplify the bending transportation of the imaging position. We can see that, because of the transportation of bending waveguide, the acoustic images formed by the hybrid sonic crystal shift 7$a$ toward left compared to the straight-line case, for both the Gaussian source and point source. Here it clearly elucidates that the image position of the hybrid sonic crystal can be designed at will by tuning the waveguide in SC1.

By now, we have shown that appropriate designing to the wave guiding function in SC1 leads to controllable imaging position of our hybrid SC. Here the main idea is based on the unique transporting ability of band gap waveguide for the acoustic waves. Based on the band gap waveguide, many interesting applications such as acoustic branching device and acoustic switch device can also be achieved. Here, referring to acoustic branching device of waveguide, we design the acoustic image branching device by the hybrid sonic crystal, as shown in Fig. 1(c). The acoustic waves radiated from the sources will be split into two branches through the waveguide in SC1, and then these two branches of acoustic waves could be focused into images respectively by the negative refraction in SC2, thus achieving acoustic image branching. In Fig. 5, we demonstrate the filed distribution of both Gaussian source and point source through such a hybrid sonic crystal. It clearly manifests the acoustic image branching as described above. Here we only illustrate the one-input-to-two-images case. By engineering branches in SC1, however, number of dividing images as well as the image positions can be designed at will, overcoming the limitation of existing SC devices that only one fixed image can be formed with one input source.

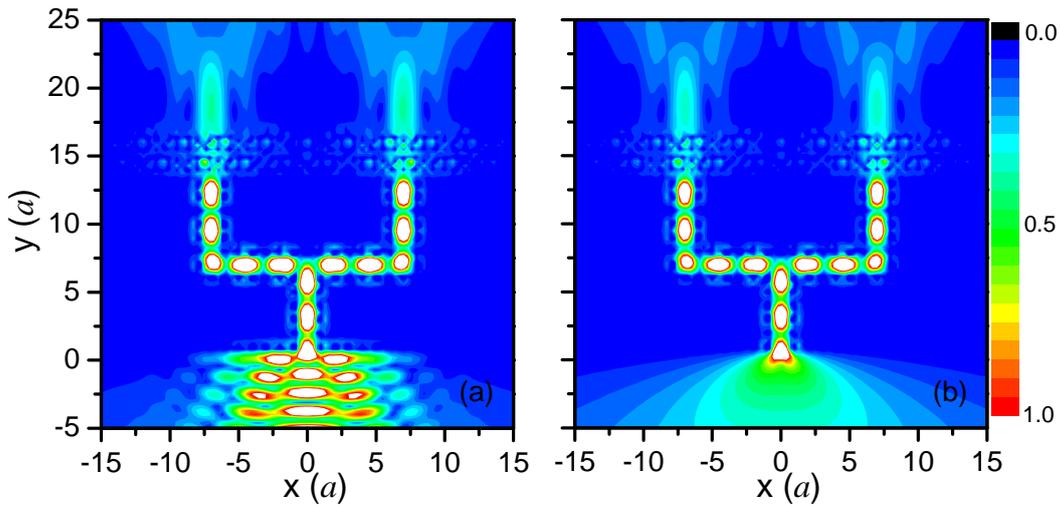

**Figure 5. Imaging pressure field of the dividing hybrid sonic crystal [structure of Fig. 1(c)] for Gaussian source incidence (a), and point source incidence (b) at frequency $0.38\omega a/2\pi c$.**

In conclusion, we have proposed a hybrid sonic crystal compositing with two simple sonic crystals to realize the acoustic image transportation and division. The main idea is based on the manipulation ability of waveguide and negative refraction for acoustic waves. We demonstrated that, as the result of combined action of wave-guiding and negative refraction, one-source-input to multi-images as well as designable image-position can be achieved by the proposed hybrid SC devices, both for the Gaussian source input and point source input. Because of the simplicity and the unique functions, our hybrid sonic crystal could have potential applications on the ultrasonic photography and compact acoustic imaging devising.

**Acknowledgment:**

This work is supported by the National Natural Science Foundation of China (Grant No. 11104113, 11264037 and 11264011), and Natural Science Foundation of Hunan province, China (Grant No. 11JJ6007), and Natural Science Foundation of Education Department of Hunan Province, China (Grant No. 11C1057).